\documentclass[aps,graphicx,12pt,showkeys,showpacs]{revtex4}
\usepackage{amsmath}
\usepackage{amscd}
\usepackage{graphicx}
\usepackage{subfigure}
\usepackage{appendix}

\begin{document}

\title[Short Title]{Fast generation of three-atom singlet state by transitionless quantum driving}

\author{Zhen Chen$^{1}$}
\author{Ye-Hong Chen$^{1}$}
\author{Yan Xia$^{1,}$\footnote{E-mail: xia-208@163.com}}
\author{Jie Song$^{2}$}
\author{Bi-Hua Huang$^{1}$}

\affiliation{$^{1}$Department of Physics, Fuzhou University, Fuzhou
350002, China\\$^{2}$Department of Physics, Harbin Institute of
Technology, Harbin 150001, China}


\begin{abstract}
Motivated by ``transitionless quantum driving'', we construct shortcuts to adiabatic passage in a three-atom system to create a
singlet state with the help of quantum zeno dynamics and non-resonant
lasers. The influence of various decoherence processes is discussed by numerical simulation and the
results reveal that the scheme is fast and robust against decoherence and operational imperfection.
We also investigate how to select the experimental parameters to control the cavity dissipation and atomic spontaneous emission which will have an application value in experiment.
\end{abstract}
\pacs {03. 67. Hk, 03. 65. Ud, 03. 67. Dd} 

\maketitle

\section*{Introduction}
Quantum entanglement is an intriguing property of composite
systems. The generation of entangled states for two
or more particles is not only fundamental for demonstrating
quantum nonlocality \cite{JSB1965,DMGMAHAJP90}, but also useful in quantum
information processing (QIP) \cite{SBZGCGPrl200,SBZAPl2009,ADEPrl91,NGSMPrl97},
typically the Bell
state, the Greenberger-Horne-Zeilinger (GHZ) state and the W state \cite{SBZEPJD2009,SYHYXJSNBAJosab2013,JSXDSQXMLLZYXPra2013,LBCMYYGWLQHDPra2007,SBZPRL2005,SBZPRA2004,SBZPRA2005}.
Different entangled state has different advantages.
For example, the W state is likely to retain bipartite entanglement when any one
of the three qubits is traced out, thus it is robust against qubit
loss. The GHZ state is the most entangled state and can maximally violate the Bell
inequalities \cite{SBZEPJD2009}. Recently, some attention has been paid to a special
type of entangled state called the $N$-particle
($N\geq2$) $N$-level singlet state \cite{ACPrl02}. The form of the
$N$-atom singlet state can be expressed as
\begin{eqnarray}\label{eq1}
|S_N\rangle=\frac{1}{\sqrt{N!}}\sum_{\{n_l\}}
{\epsilon_{n_1,n_2,\cdots,n_N}|g_{n_1},g_{n_2},\cdots,g_{n_N}\rangle},
\end{eqnarray}
where $\epsilon_{n_1,n_2,\cdots,n_N}$ are the generalized
Levi-Civita symbols, $\{n_{l}\}$ are the permutations, and
$|g_{n_l}\rangle$ denote the bases of the qubits \cite{YHCQip13}. It
has been shown that the singlet state not only is in connection with
violations of Bell inequalities \cite{NDMPrd80}, but also can be
used to construct decoherence-free subspace, which is robust against
collective decoherence \cite{ACJmo03}. Moreover, the singlet state
can be used to solve several problems which have no classical
solutions, including ``$N$ strangers'', ``secret sharing'', ``liar
detection'', and so on \cite{ACJmo03,ACPrl02}. Furthermore, the
singlet state also can be used in a scheme designed to probe a
quantum gate that can realize an unknown unitary transformation
\cite{MHVBPra01}. In recent years, lots of theoretical schemes have been proposed to generate
singlet state in the cavity quantum electrodynamics (C-QED) system via different techniques
\cite{ACJmo03,MHVBPra01,GSJSSLSLFHZZPra05,GWLMYYLBCQHDXMLPra05,XQSHFWLCSZYFZKHYNjp10,ZCSYXJSHSSQip13,MLYXJSHSSJpb13}.
Among these techniques \cite{ACJmo03,MHVBPra01,GSJSSLSLFHZZPra05,GWLMYYLBCQHDXMLPra05,XQSHFWLCSZYFZKHYNjp10,ZCSYXJSHSSQip13,MLYXJSHSSJpb13},
there are two famous techniques for their robustness against
decoherence in proper conditions.
One is stimulated Raman adiabatic passage (STIRAP) \cite{GWLMYYLBCQHDXMLPra05,MLYXJSHSSJpb13},
the other is Quantum Zeno dynamics (QZD) \cite{YHCQip13,XQSHFWLCSZYFZKHYNjp10,ZCSYXJSHSSQip13}.
In general, adiabatic passage technique has been widely used and
an advantage of the technique is that can reduce populations of the intermediate excited states.
Therefore, the technique would restrain the influence of atomic spontaneous emission on the fidelity.
As we know, the adiabatic condition is managed to be slow to make sure
each of the eigenstates of the system evolves along itself all the time without transition to other ones.
So, the operation time is long in previous schemes \cite{GWLMYYLBCQHDXMLPra05,MLYXJSHSSJpb13} via adiabatic passage.
Differ from the adiabatic passage, QZD is usually robust against photon leakage but sensitive to atomic spontaneous emission \cite{YHCQip13,XQSHFWLCSZYFZKHYNjp10,ZCSYXJSHSSQip13}.
Therefore, some of the researchers introduce detuning
between the atomic transition to restrain the influence of atomic
spontaneous emission. However, that also
increases the operation time. In general, the interaction time for
a method is the shorter the better. Otherwise, the method
may be useless because the dissipation caused by decoherence,
noise, and losses on the target state increases with the increasing
of the interaction time \cite{YHCYXQQCJS}.

In order to solve this problem, in recent years researchers pay more attention to
``shortcuts to adiabatic passage (STAP)'' \cite{XCILARDGJGMPrl10,ETSISMMMADAXC13,SMKNPra2008,MMLJNKTCPra2009} which employs
a set of techniques to speed up a slow quantum adiabatic
process through a non-adiabatic route. Usually STAP can overcome
the harmful effect caused by decoherence, noise and losses during
the long operation time.
Recently, STAP has been applied in a wide range of system to implement quantum information processing (QIP) in theory and experiment \cite{XCILARDGJGMPrl10,ETSISMMMADAXC13,SMKNPra2008,MMLJNKTCPra2009,XCETJGMPra11,JGMXCARDGJpb09,HRLWBRJmp1969,MVBJpa2009,MGBMVNMPHNp1012,MDSARJpca2003,MDSARJcP2008,ADPrl2013,SMKNPra2011,YLQCWSLSXJSZPra15,ADCPrA2011,ADCMGBSCIR2012,JFSPCGLPVNJP2011,YHCYXQQCJSPra14,MSNKDCPrl14,DSJCDAPrX14,DCAMJGKMPrA08,MLYXLTSJSNBAPra14,YHCYXQQCJS2014,MLYXLTSJSLp2014,YHCYXQQCJSLpl14,YHCYXQQCJSPra2015,JGMXCARDGJpb2009,XCARSSADDJGMPrl2010,XCJGMPra2010,DSJRJSLPra2010,ATRPrl2009,ETSMARJGMPra2012,SCROBSPra2011}.
In order to construct STAP to speed up adiabatic processes effectively,
many methods \cite{XCILARDGJGMPrl10,ETSISMMMADAXC13,SMKNPra2008,MMLJNKTCPra2009,XCETJGMPra11,JGMXCARDGJpb09,HRLWBRJmp1969,MVBJpa2009,MGBMVNMPHNp1012,MDSARJpca2003,MDSARJcP2008,ADPrl2013,SMKNPra2011,YLQCWSLSXJSZPra15,ADCPrA2011,ADCMGBSCIR2012,JFSPCGLPVNJP2011} are related.
Such as, invariant-based
inverse engineering proposed by Muga and Chen \cite{XCILARDGJGMPrl10,ETSISMMMADAXC13,SMKNPra2008,MMLJNKTCPra2009,XCETJGMPra11,JGMXCARDGJpb09,HRLWBRJmp1969},
can achieve the fast population transfer within two internal states of a single $\Lambda$-type atom.
``Transitionless quantum
driving'' (TQD) \cite{MVBJpa2009,MGBMVNMPHNp1012,MDSARJpca2003,MDSARJcP2008} proposed by Berry,
provides a very effective method to construct the ``counter-diabatic driving'' (CDD)
Hamiltonian $H(t)$ which can accurately derive the instantaneous
eigenstates of $H_{0}(t)$ to speed up adiabatic processes effectively.
But it is also found that the designed
CDD Hamiltonian is hard to be directly implemented
in practice, especially in multiparticle system.
In order to solve the problem, many schemes \cite{XCETJGMPra11,MGBMVNMPHNp1012,MDSARJpca2003,MLYXLTSJSNBAPra14,YHCYXQQCJS2014} have been put forward.
In 2014, by using second-order perturbation
approximation twice under large detuning condition
and transitionless quantum driving, Lu \emph{et al.} have proposed an
effective scheme \cite{MLYXLTSJSNBAPra14} to implement the fast populations transfer
and prepare a fast maximum entanglement between two
atoms in a cavity. The idea inspires that using some
traditional methods to approximate a complicated Hamiltonian
into an effective and simple one first, then constructing
shortcuts for the effective Hamiltonian might be a promising
method to speed up evolution process of a system. Later,
Chen \emph{et al.} \cite{YHCYXQQCJS2014} have proposed a promising method to construct STAP for a three-atom system to
generate GHZ states in the cavity QED system in light of QZD and TQD.
Their schemes might be useful to realize fast and noise-resistant
quantum information processing for multi-qubit system in current technology.

In this paper inspired by the schemes \cite{MLYXLTSJSNBAPra14,YHCYXQQCJS2014}, we discuss how to construct
STAP to fastly generate a three-atom singlet state in cavity QED system by using the approach of ``transitionless tracking algorithm''.
Based on quantum Zeno dynamics \cite{PKHWTHAZMAPrl1995,PFSPPrl2002}
and large detuning conditon, we can simplify the original Hamiltonian of system and obtain the
effective Hamiltonian equivalent to the corresponding
CDD Hamiltonian, the evolution process of system can be speeded up, and the STAP can be achieved in experiment easily.
What's more, numerical investigation shows that our scheme is also fast and robust against
both cavity decay and atomic spontaneous emission for three-atom singlet state preparation.
It will be much useful in dealing with the
fast and noise-resistant generation of $N$-atom singlet state.

The paper is organized as follows. In section \textrm{II}, we
describe a theoretical model for three atoms which are trapped in a
bimodal-mode cavity.
In section \textrm{III}, we demonstrate how to construct STAP for the system in section \textrm{II}, and use
the constructed shortcut to generate a three-atom singlet state.
The numerical simulation and experimental discussion
about the validity of the scheme are also given.
Finally, a summary is given in
section \textrm{IV}.

\section*{Theoretical model}

The sketch of the experimental setup is shown in figure 1.
Three identical four-level
atoms with three ground states $|g_{0}\rangle$, $|g_{1}\rangle$ and $|g_{2}\rangle$, and an excited state
$|e\rangle$ are trapped in a bimodal-mode cavity.
The atomic transition $|g_{2}\rangle\leftrightarrow|e\rangle$ is driven resonantly through
classical laser field with time-dependent Rabi frequency $\Omega(t)$, the transition $|g_{0}\rangle\leftrightarrow|e\rangle$
is coupled resonantly to the left-circularly polarized mode of the cavity with coupling $\lambda_{L}$,
and transition $|g_{1}\rangle\leftrightarrow|e\rangle$ is coupled resonantly to the right-circularly polarized
mode of the cavity with coupling $\lambda_{R}$. Under the rotating-wave approximation (RWA),
the interaction Hamiltonian for this system reads ($\hbar=1$):
\begin{eqnarray}\label{eq2}
H_{I}&=&H_{al}+H_{ac},         \cr\cr
H_{al}&=&\sum_{i=1,2,3}\Omega_{i}(t)|e\rangle_{i}\langle g_{2}|+H.c.,    \cr\cr
H_{ac}&=&\sum_{i=1,2,3}(\lambda_{L,i}|e\rangle_{i}\langle g_{0}|a_{L}+\lambda_{R,i}|e\rangle_{i}\langle g_{1}|a_{R})+H.c.,
\end{eqnarray}
where $a_{L}$ and $a_{R}$ are the left-circularly and the right-circularly annihilation operators for cavity mode, respectively.
We set $\lambda_{L,i}=\lambda_{R,i}=\lambda$ for simplicity. If we assume the initial state of the system is
$-\frac{1}{\sqrt{2}}(|g_{2},g_{0},g_{1}\rangle_{1,2,3}|00\rangle_{a_{L},a_{R}}-|g_{2},g_{1},g_{0}\rangle_{1,2,3}|00\rangle_{a_{L},a_{R}})$,
the system will evolve within a single-excitation subspace with basis states
\begin{eqnarray}\label{eq3}
|\phi_{1}\rangle&=&|g_{2},g_{0},g_{1}\rangle_{1,2,3}|00\rangle_{a_{L},a_{R}}, \ \ \
|\phi_{2}\rangle=|e,g_{0},g_{1}\rangle_{1,2,3}|00\rangle_{a_{L},a_{R}},    \cr
|\phi_{3}\rangle&=&|g_{0},g_{0},g_{1}\rangle_{1,2,3}|10\rangle_{a_{L},a_{R}}, \ \ \
|\phi_{4}\rangle=|g_{0},e,g_{1}\rangle_{1,2,3}|00\rangle_{a_{L},a_{R}},    \cr
|\phi_{5}\rangle&=&|g_{0},g_{1},g_{1}\rangle_{1,2,3}|01\rangle_{a_{L},a_{R}}, \ \ \
|\phi_{6}\rangle=|g_{0},g_{1},e\rangle_{1,2,3}|00\rangle_{a_{L},a_{R}},    \cr
|\phi_{7}\rangle&=&|g_{0},g_{1},g_{0}\rangle_{1,2,3}|10\rangle_{a_{L},a_{R}}, \ \ \
|\phi_{8}\rangle=|g_{1},g_{0},g_{1}\rangle_{1,2,3}|01\rangle_{a_{L},a_{R}},    \cr
|\phi_{9}\rangle&=&|g_{1},g_{0},e\rangle_{1,2,3}|00\rangle_{a_{L},a_{R}}, \ \ \
|\phi_{10}\rangle=|g_{1},g_{0},g_{0}\rangle_{1,2,3}|10\rangle_{a_{L},a_{R}},    \cr
|\phi_{11}\rangle&=&|g_{1},e,g_{0}\rangle_{1,2,3}|00\rangle_{a_{L},a_{R}}, \ \ \
|\phi_{12}\rangle=|g_{1},g_{1},g_{0}\rangle_{1,2,3}|01\rangle_{a_{L},a_{R}},    \cr
|\phi_{13}\rangle&=&|e,g_{1},g_{0}\rangle_{1,2,3}|00\rangle_{a_{L},a_{R}}, \ \ \
|\phi_{14}\rangle=|g_{2},g_{1},g_{0}\rangle_{1,2,3}|00\rangle_{a_{L},a_{R}},    \cr
|\phi_{15}\rangle&=&|g_{0},g_{2},g_{1}\rangle_{1,2,3}|00\rangle_{a_{L},a_{R}}, \ \ \
|\phi_{16}\rangle=|g_{0},g_{1},g_{2}\rangle_{1,2,3}|00\rangle_{a_{L},a_{R}},    \cr
|\phi_{17}\rangle&=&|g_{1},g_{0},g_{2}\rangle_{1,2,3}|00\rangle_{a_{L},a_{R}}, \ \ \
|\phi_{18}\rangle=|g_{1},g_{2},g_{0}\rangle_{1,2,3}|00\rangle_{a_{L},a_{R}}.
\end{eqnarray}
Then, we rewrite the Hamiltonian $H_{ac}$ and $H_{al}$ with the eigenvectors of $H_{ac}$:
\begin{eqnarray}\label{eq4}
|\Psi_{1}\rangle&=&\frac{1}{\sqrt{6}}(-|\phi_{3}\rangle+|\phi_{5}\rangle-|\phi_{7}\rangle+|\phi_{8}\rangle-|\phi_{10}\rangle+|\phi_{12}\rangle),\cr
|\Psi_{2}\rangle&=&\frac{1}{\sqrt{6}}(-|\phi_{2}\rangle+|\phi_{4}\rangle-|\phi_{6}\rangle+|\phi_{9}\rangle-|\phi_{11}\rangle+|\phi_{13}\rangle),\cr
|\Psi_{3}\rangle&=&\frac{1}{2\sqrt{2}}(|\phi_{3}\rangle+|\phi_{4}\rangle-|\phi_{6}\rangle-|\phi_{7}\rangle-|\phi_{8}\rangle-|\phi_{9}\rangle+|\phi_{11}\rangle+|\phi_{12}\rangle),\cr
|\Psi_{4}\rangle&=&\frac{1}{2\sqrt{2}}(|\phi_{2}\rangle-|\phi_{4}\rangle-|\phi_{5}\rangle+|\phi_{7}\rangle+|\phi_{8}\rangle-|\phi_{10}\rangle-|\phi_{11}\rangle+|\phi_{13}\rangle),\cr
|\Psi_{5}\rangle&=&\frac{1}{2\sqrt{2}}(|\phi_{3}\rangle-|\phi_{4}\rangle+|\phi_{6}\rangle-|\phi_{7}\rangle-|\phi_{8}\rangle+|\phi_{9}\rangle-|\phi_{11}\rangle+|\phi_{12}\rangle),\cr
|\Psi_{6}\rangle&=&\frac{1}{2\sqrt{2}}(|\phi_{2}\rangle-|\phi_{4}\rangle+|\phi_{5}\rangle-|\phi_{7}\rangle-|\phi_{8}\rangle+|\phi_{10}\rangle-|\phi_{11}\rangle+|\phi_{13}\rangle),\cr
|\Psi_{7}\rangle&=&\frac{1}{2\sqrt{3}}(|\phi_{2}\rangle+|\phi_{3}\rangle+|\phi_{4}\rangle+|\phi_{5}\rangle+|\phi_{6}\rangle+|\phi_{7}\rangle+|\phi_{8}\rangle+|\phi_{9}\rangle \cr
&+&|\phi_{10}\rangle+|\phi_{11}\rangle+|\phi_{12}\rangle+|\phi_{13}\rangle),\cr
|\Psi_{8}\rangle&=&\frac{1}{2\sqrt{3}}(|\phi_{2}\rangle-|\phi_{3}\rangle+|\phi_{4}\rangle-|\phi_{5}\rangle+|\phi_{6}\rangle-|\phi_{7}\rangle-|\phi_{8}\rangle+|\phi_{9}\rangle \cr
&-&|\phi_{10}\rangle+|\phi_{11}\rangle-|\phi_{12}\rangle+|\phi_{13}\rangle),\cr
|\Psi_{9}\rangle&=&\frac{1}{2\sqrt{6}}(-|\phi_{3}\rangle-\sqrt{3}|\phi_{4}\rangle-2|\phi_{5}\rangle-\sqrt{3}|\phi_{6}\rangle-|\phi_{7}\rangle+|\phi_{8}\rangle+\sqrt{3}|\phi_{9}\rangle \cr
&+&2|\phi_{10}\rangle+\sqrt{3}|\phi_{11}\rangle+|\phi_{12}\rangle),\cr
|\Psi_{10}\rangle&=&\frac{1}{2\sqrt{6}}(-|\phi_{2}\rangle+|\phi_{4}\rangle+\sqrt{3}|\phi_{5}\rangle+2|\phi_{6}\rangle+\sqrt{3}|\phi_{7}\rangle-\sqrt{3}|\phi_{8}\rangle-2|\phi_{9}\rangle \cr
&-&\sqrt{3}|\phi_{10}\rangle-|\phi_{11}\rangle+|\phi_{13}\rangle),\cr
|\Psi_{11}\rangle&=&\frac{1}{2\sqrt{6}}(-|\phi_{3}\rangle+\sqrt{3}|\phi_{4}\rangle-2|\phi_{5}\rangle+\sqrt{3}|\phi_{6}\rangle-|\phi_{7}\rangle+|\phi_{8}\rangle-\sqrt{3}|\phi_{9}\rangle \cr
&+&2|\phi_{10}\rangle-\sqrt{3}|\phi_{11}\rangle+|\phi_{12}\rangle),\cr
|\Psi_{12}\rangle&=&\frac{1}{2\sqrt{6}}(-|\phi_{2}\rangle+|\phi_{4}\rangle-\sqrt{3}|\phi_{5}\rangle+2|\phi_{6}\rangle-\sqrt{3}|\phi_{7}\rangle+\sqrt{3}|\phi_{8}\rangle-2|\phi_{9}\rangle \cr
&+&\sqrt{3}|\phi_{10}\rangle-|\phi_{11}\rangle+|\phi_{13}\rangle),
\end{eqnarray}
with eigenvalues $\eta_{1}=\eta_{2}=0$, $\eta_{3}=\eta_{4}=\lambda$, $\eta_{5}=\eta_{6}=-\lambda$, $\eta_{7}=2\lambda$, $\eta_{8}=-2\lambda$, $\eta_{9}=\eta_{10}=\sqrt{3}\lambda$,
and $\eta_{11}=\eta_{12}=-\sqrt{3}\lambda$. We obtain
\begin{eqnarray}\label{eq5}
H_{ac}^{\prime}&=&\sum_{n=1}^{12}\eta_{n}|\Psi_{n}\rangle\langle\Psi_{n}|, \cr\cr
H_{al}^{\prime}&=&\frac{1}{\sqrt{6}}|\Psi_{2}\rangle(-\Omega_{1}\langle\phi_{1}|+\Omega_{1}\langle\phi_{14}|+\Omega_{2}\langle\phi_{15}|-\Omega_{3}\langle\phi_{16}|+\Omega_{3}\langle\phi_{17}|-\Omega_{2}\langle\phi_{18}|)\cr
&+&\frac{1}{2\sqrt{2}}(|\Psi_{3}\rangle-|\Psi_{5}\rangle)(\Omega_{2}\langle\phi_{15}|-\Omega_{3}\langle\phi_{16}|+\Omega_{3}\langle\phi_{17}|+\Omega_{2}\langle\phi_{18}|)\cr
&+&\frac{1}{2\sqrt{2}}(|\Psi_{4}\rangle+|\Psi_{6}\rangle)(\Omega_{1}\langle\phi_{1}|+\Omega_{1}\langle\phi_{14}|-\Omega_{2}\langle\phi_{15}|-\Omega_{2}\langle\phi_{18}|)\cr
&+&\frac{1}{2\sqrt{3}}(|\Psi_{7}\rangle+|\Psi_{8}\rangle)(\Omega_{1}\langle\phi_{1}|+\Omega_{1}\langle\phi_{14}|+\Omega_{2}\langle\phi_{15}|+\Omega_{3}\langle\phi_{16}|+\Omega_{3}\langle\phi_{17}|+\Omega_{2}\langle\phi_{18}|)\cr
&+&\frac{1}{2\sqrt{6}}(|\Psi_{10}\rangle+|\Psi_{12}\rangle)(-\Omega_{1}\langle\phi_{1}|+\Omega_{1}\langle\phi_{14}|+\Omega_{2}\langle\phi_{15}|+2\Omega_{3}\langle\phi_{16}|-2\Omega_{3}\langle\phi_{17}|-\Omega_{2}\langle\phi_{18}|)\cr
&+&\frac{1}{2\sqrt{2}}(|\Psi_{9}\rangle-|\Psi_{11}\rangle)(-\Omega_{2}\langle\phi_{15}|-\Omega_{3}\langle\phi_{16}|+\Omega_{3}\langle\phi_{17}|+\Omega_{2}\langle\phi_{18}|)+H.c..
\end{eqnarray}
Through performing the unitary transformation $U=\exp(-iH_{ac}^{\prime}t)$ and neglecting the terms with high oscillating
frequency by setting the condition $\Omega_{i}\ll\lambda$, we obtain an effective Hamiltonian
\begin{eqnarray}\label{eq6}
H_{eff}=\frac{1}{\sqrt{3}}\Omega_{1}|\Psi_{2}\rangle\langle\chi|+\frac{2}{\sqrt{6}}\Omega_{3}|\Psi_{2}\rangle\langle\varpi|+H.c.,
\end{eqnarray}
here we set $\Omega_{2}=\Omega_{3}$, $|\chi\rangle=\frac{1}{\sqrt{2}}(-|\phi_{1}\rangle+|\phi_{14}\rangle)$ and
$|\varpi\rangle=\frac{1}{2}(|\phi_{15}\rangle-|\phi_{16}\rangle+|\phi_{17}\rangle-|\phi_{18}\rangle)$.

We can see Hamiltonian in equation (\ref{eq6}) as a simple three-level system with an excited
state $|\Psi_{2}\rangle$ and two ground states $|\chi\rangle$ and $|\varpi\rangle$.
For this effective Hamiltonian, its eigenstates are easily obtained
\begin{eqnarray}\label{eq7}
|n_{0}(t)\rangle=\left(\begin{array}{c}
                         -\cos\theta(t)\\
                         0\\
                         \sin\theta(t)
                       \end{array}
\right), |n_{\pm}(t)\rangle=\frac{1}{\sqrt{2}}\left(\begin{array}{c}
                         \sin\theta(t)\\
                         \pm1\\
                         \cos\theta(t)
                       \end{array}
\right),
\end{eqnarray}
corresponding eigenvalues $\varepsilon_{0}=0$, $\varepsilon_{\pm}=\pm\frac{\Omega}{\sqrt{3}}$, respectively,
where $\tan\theta=\frac{\Omega_{1}}{\sqrt{2}\Omega_{3}}$ and $\Omega=\sqrt{\Omega_{1}^{2}+2\Omega_{3}^{2}}$.
When the adiabatic condition $|\langle n_{0}|\partial_{t}n_{\pm}\rangle|\ll|\varepsilon_{\pm}|$ is fulfilled, the initial state
$|\psi_{1}\rangle=-\frac{1}{\sqrt{2}}(|g_{2},g_{0},g_{1}\rangle_{1,2,3}|00\rangle_{a_{L},a_{R}}-|g_{2},g_{1},g_{0}\rangle_{1,2,3}|00\rangle_{a_{L},a_{R}})
=\frac{1}{\sqrt{2}}(-|\phi_{1}\rangle+|\phi_{14}\rangle)=|\chi\rangle=|n_{0}(0)\rangle$ will follow $|n_{0}(t)\rangle$ closely,
and when $\tan\theta(t)=\sqrt{2}$, we can obtain the three-atom singlet state:
\begin{eqnarray}\label{eq8}
|\psi(t_{f})\rangle&=&-\frac{1}{\sqrt{3}}|\chi\rangle+\frac{\sqrt{2}}{\sqrt{3}}|\varpi\rangle \cr\cr
&=&\frac{1}{\sqrt{6}}(-|\phi_{1}\rangle+|\phi_{14}\rangle+|\phi_{15}\rangle-|\phi_{16}\rangle+|\phi_{17}\rangle-|\phi_{18}\rangle)  \cr\cr
&=&\frac{1}{\sqrt{6}}(-|g_{2}g_{0}g_{1}\rangle+|g_{2}g_{1}g_{0}\rangle+|g_{0}g_{2}g_{1}\rangle-|g_{0}g_{1}g_{2}\rangle   \cr\cr
&+&|g_{1}g_{0}g_{2}\rangle-|g_{1}g_{2}g_{0}\rangle)\otimes|00\rangle.
\end{eqnarray}
However, this process will take quite a long time to obtain
the target state, which is undesirable. We will talk in later.

\section*{Using STAP to generate a three-atom singlet state}
The instantaneous eigenstates $|n_{k}\rangle$ ($k=0,\pm$) for the effective Hamiltonian $H_{eff}(t)$
in equation (\ref{eq6}) do not satisfy the Schrodinger equation $i\partial_{t}|n_{k}\rangle=H_{eff}(t)|n_{k}\rangle$.
According to Berry's transitionless tracking algorithm \cite{MVBJpa2009}, from $H_{eff}(t)$, one can reverse engineer
$H(t)$ which is related to the original Hamiltonian $H_{I}(t)$ and can drive the eigenstates exactly.
From refs. \cite{MLYXLTSJSNBAPra14,XCARSSADDJGMPrl2010,XCJGMPra2010}, we learn the simplest Hamiltonian $H(t)$ is derived in the form
\begin{eqnarray}\label{eq9}
H(t)=i\sum_{k=0,\pm}|\partial_{t}n_{k}(t)\rangle\langle n_{k}(t)|.
\end{eqnarray}
Substituting equation (\ref{eq7}) into equation (\ref{eq9}), we obtain
\begin{eqnarray}\label{eq10}
H(t)=i\dot{\theta}|\chi\rangle\langle\varpi|+H.c.,
\end{eqnarray}
where $\dot{\theta}(t)=[\sqrt{2}(\dot{\Omega_{1}}(t)\Omega_{3}(t)-\dot{\Omega_{3}}(t)\Omega_{1}(t))]/\Omega^{2}$.
For this three-atom system, the Hamiltonian $H(t)$
is hard or even impossible to be implemented in real experiment \cite{MLYXLTSJSNBAPra14}. We should find an alternative physically feasible (APF) Hamiltonian
whose effect is equivalent to $H(t)$. Therefore, we consider that the three atoms are trapped in a cavity and the atomic level configuration is shown in figure 2.
We make all the resonant atomic transitions into non-resonant atomic
transitions with detuning $\Delta$. The non-resonant Hamiltonian
reads
\begin{eqnarray}\label{eq11}
H_{I}^{\prime}&=&H_{al}^{\prime}+H_{ac}^{\prime}+H_{e},         \cr\cr
H_{al}^{\prime}&=&\sum_{i=1,2,3}\Omega_{i}^{\prime}(t)|e\rangle_{i}\langle g_{2}|+H.c.,    \cr\cr
H_{ac}^{\prime}&=&\sum_{i=1,2,3}(\lambda_{L,i}|e\rangle_{i}\langle g_{0}|a_{L}+\lambda_{R,i}|e\rangle_{i}\langle g_{1}|a_{R})+H.c.,    \cr\cr
H_{e}&=&\sum_{i=1,2,3}\Delta|e\rangle_{i}\langle e|.
\end{eqnarray}
Then similar to the approximation for the Hamiltonian from equation (\ref{eq2}) to
equation (\ref{eq6}) in section \textrm{II}, we also obtain an effective Hamiltonian
for the present non-resonant system \cite{YHCQip13}
\begin{eqnarray}\label{eq12}
H_{eff}^{\prime}=(\frac{1}{\sqrt{3}}\Omega_{1}^{\prime}|\Psi_{2}\rangle\langle\chi|+\frac{2}{\sqrt{6}}\Omega_{3}^{\prime}|\Psi_{2}\rangle\langle\varpi|+H.c.)
+\Delta|\Psi_{2}\rangle\langle\Psi_{2}|.
\end{eqnarray}
By adiabatically eliminating the state $|\Psi_{2}\rangle$ under the condition $\Delta\gg\Omega_{1}^{\prime},\Omega_{3}^{\prime}$,
we obtain the final effective Hamiltonian
\begin{eqnarray}\label{eq13}
H_{fe}=\frac{\Omega_{1}^{\prime2}}{3\Delta}|\chi\rangle\langle\chi|+\frac{2\Omega_{3}^{\prime2}}{3\Delta}|\varpi\rangle\langle\varpi|
+(\frac{\sqrt{2}\Omega_{1}^{\prime}\Omega_{3}^{\prime}}{3\Delta}|\chi\rangle\langle\varpi|+H.c.).
\end{eqnarray}
When we choose $\Omega_{1}^{\prime}=\Omega^{\prime}$ and $\Omega_{3}^{\prime}=i\Omega^{\prime}/\sqrt{2}$ (here $\Omega^{\prime}$ is a real number),
the first two terms can be removed, and the Hamiltonian in equation (\ref{eq13}) becomes
\begin{eqnarray}\label{eq14}
\widetilde{H}_{eff}=i\frac{\Omega^{\prime2}}{3\Delta}|\chi\rangle\langle\varpi|+H.c..
\end{eqnarray}
That means, as long as $\frac{\Omega^{\prime2}}{3\Delta}=\dot{\theta}$, $\widetilde{H}_{eff}=H_{CDD}$,
the Hamiltonian for speeding up the adiabatic dark state evolution governed
by $H_{I}^{\prime}$ under condition $\Omega_{1}^{\prime}, \Omega_{3}^{\prime}\ll\lambda, \Delta$ has been constructed.
Hence, $\Omega^{\prime}$ is given
\begin{eqnarray}\label{eq15}
\Omega^{\prime}=\sqrt{3\Delta\dot{\theta}}=\sqrt{\frac{3\sqrt{2}\Delta(\dot{\Omega}_{1}\Omega_{3}-\Omega_{1}\dot{\Omega}_{3})}{\Omega^{2}}}.
\end{eqnarray}

We will show the numerical analysis of the creation of a three-atom singlet state governed by $H_{I}^{\prime}$.
To satisfy the boundary condition of the fractional stimulated
Raman adiabatic passage (STIRAP),
\begin{eqnarray}\label{eq16}
\lim_{t\rightarrow-\infty}\frac{\Omega_{1}(t)}{\Omega_{3}(t)}=0, \lim_{t\rightarrow+\infty}\frac{\Omega_{1}(t)}{\Omega_{3}(t)}=\tan\alpha,
\end{eqnarray}
the Rabi frequencies $\Omega_{1}(t)$ and $\Omega_{3}(t)$ in the original Hamiltonian $H_{I}(t)$ are chosen as
\begin{eqnarray}\label{eq17}
\Omega_{1}(t)&=&\sin\alpha\Omega_{0}\exp^{-(t-t_{f}/2-t_{0})^{2}/t_{c}^{2}},      \cr\cr
\Omega_{3}(t)&=&\cos\alpha\Omega_{0}\exp^{-(t-t_{f}/2-t_{0})^{2}/t_{c}^{2}}+\Omega_{0}\exp^{-(t-t_{f}/2+t_{0})^{2}/t_{c}^{2}},
\end{eqnarray}
where $\Omega_{0}$ is the pulse amplitude, $t_{f}$ is the operation time, and $t_{0}$, $t_{c}$ are some related parameters.
In order to create a three-atom singlet state, the finial state $|\psi(t_{f})\rangle$ should be
$|\psi(t_{f})\rangle=\frac{1}{\sqrt{6}}(-|\phi_{1}\rangle+|\phi_{14}\rangle+|\phi_{15}\rangle-|\phi_{16}\rangle+|\phi_{17}\rangle-|\phi_{18}\rangle)$
according to equation (\ref{eq8}).
Therefore, we have $\tan\alpha=2$. And choosing parameters for the
laser pulses suitably to fulfill the boundary condition in equation (\ref{eq16}),
the time-dependent $\Omega_{1}(t)$ and $\Omega_{3}(t)$ are gotten as shown
in figure 3 with parameters $t_{0}=0.14t_{f}$ and $t_{c}=0.19t_{f}$.

Figure 4 shows the relationship between the fidelity of the generated
three-atom singlet state (governed by the APF Hamiltonian $H_{I}^{\prime}(t)$) and two parameters
$\Delta$ and $t_{f}$ when $\Omega_{0}=0.2\lambda$, where the fidelity for the three-atom
singlet state is given through $F=|\langle S|\rho(t_{f})|S\rangle|$ ($\rho(t_{f})$ is
the density operator of the whole system when $t=t_{f}$). It's easy to find that there is a wide range of selectable values for
parameters $\Delta$ and $t_{f}$ to get a high fidelity.
And the fidelity increases with the increasing of $t_{f}$ while decreases with the increasing of $\Delta$.
This is easy to understand.
If we set $t^{\prime}=\frac{t}{t_{f}}$, according to equation (\ref{eq17}), we can obtain two
dimensionless parameters
\begin{eqnarray}\label{eq18}
x&=&\frac{t^{\prime}t_{f}-t_{0}-0.5t_{f}}{t_{c}},         \cr\cr
y&=&\frac{t^{\prime}t_{f}+t_{0}-0.5t_{f}}{t_{c}}.
\end{eqnarray}
Therefore, putting equations (\ref{eq17}) and (\ref{eq18}) into equation (\ref{eq15}), we obtain
\begin{eqnarray}\label{eq19}
\Omega^{\prime}=\sqrt{\frac{6\sqrt{2}\Delta}{t_{f}}G^{2}},
\end{eqnarray}
where
\begin{eqnarray}\label{eq20}
G^{2}=|\frac{-\Omega_{1}\Omega_{0}(e^{-y^{2}}x-e^{-y^{2}}y)}{\Omega_{1}^{2}+2\Omega_{3}^{2}}|,
\end{eqnarray}
is a dimensionless function. A brief analysis of $G$ tells
that the amplitude of $G$ is close to 1. That is, the amplitude
of $\Omega^{\prime}$ is mainly dominated by $\sqrt{\frac{6\sqrt{2}\Delta}{t_{f}}}$.
In order to satisfy the condition $\Omega^{\prime}\ll\lambda$ and $\Omega^{\prime}\ll\Delta$,
we can work out $\Delta/t_{f}\ll1$ and $\Delta t_{f}\gg1$. So, long $t_{f}$ can lead to a high fidelity
as shown in figure 4. When the detuning $\Delta$ is smaller or near 0, it's not meet the condition $\Delta t_{f}\gg1$,
so the fidelity is lower in a short time as shown in figure 4.
We know $\Omega^{\prime}\approx\sqrt{\frac{6\sqrt{2}\Delta}{t_{f}}}$,
shortening the evolution time implies that relative large
laser intensities is required, and this would destroy the Zeno condition.
Yet slightly destroying the Zeno condition is also helpful to achieve the target
state in a much shorter interaction time \cite{MLYXLTSJSNBAPra14,YHCYXQQCJSPra14}.

Next, to comfirm the operation time required for the creation of the
three-atom singlet state governed by $H_{I}^{\prime}$ is much shorter than
that governed by $H_{I}$, we contrast the performances of population
transfer from the initial state $|\psi_{1}\rangle$
in figure 5. The time-dependent population for any state $|\psi\rangle$ is given by
$P=|\langle\psi|\rho(t)|\psi\rangle|$, where $\rho(t)$ is the corresponding
time-dependent density operator. Figure 5(a) shows time evolution of the populations for the states
$|\chi\rangle$ ($|\chi\rangle$ is the initial state $|\psi_{1}\rangle$) and $|\varpi\rangle$ governed by the APF Hamiltonian $H_{I}^{\prime}$
with $\Omega_{0}=0.2\lambda$, $t_{f}=40/\lambda$ and $\Delta=3\lambda$.
Figure 5(b) shows time evolution of the populations for the states
$|\chi\rangle$ and $|\varpi\rangle$ governed by the original Hamiltonian $H_{I}$
with $\Omega_{0}=0.2\lambda$ and $t_{f}=1000/\lambda$.
The comparison of figure 5 (a) and (b) shows that with this set of parameters,
the APF Hamiltonian $H_{I}^{\prime}$ can govern the evolution
to achieve a near-perfect three-atom singlet state from state $|\psi_{1}\rangle$
in short interaction time while the original Hamiltonian $H_{I}$ can not.
We also plot the fidelities of the evolved states governed by $H_{I}^{\prime}$ and $H_{I}$
in figure 6, with respect to the target three-atom singlet state.
As shown in figure 6, when the interaction time $t_{f}=40/\lambda$, the fidelity
governed by $H_{I}^{\prime}$ is already 99.98\%. While, when $t_{f}=1000/\lambda$, the fidelity
governed by $H_{I}$ achieves 99.93\%. The interaction time required for creation of
the three-atom singlet state via STAP is much shorter than adiabatic passage.

Since most of the parameters are hard to faultlessly achieve in experiment,
we need to investigate the variations in the parameters induced by the experimental imperfection.
We calculate the fidelity by varying error parameters of the mismatch between the laser amplitude
$\Omega_{0}$ and the total operation time $t_{f}$, the detuning $\Delta$ and
the cavity mode with coupling constant $\lambda$, respectively.
We define $\delta x=x'-x$ as the deviation of $x$, here $x$ denotes
the ideal value and $x'$ denotes the actual value.
Then in figure 7(a) we plot the fidelity of the three-atom singlet state
versus the variations in total operation time $t_{f}$ and laser amplitude $\Omega_{0}$.
In figure 7(b) we plot the fidelity of the three-atom singlet state
versus the variations in coupling constant $\lambda$ and the detuning $\Delta$.
We find that the scheme is robust against all of these variations.
For example, a deviation $\delta \Delta/\Delta=10\%$ and $\delta
\lambda/\lambda=-10\%$ only causes a reduction about $0.66\%$ in the fidelity.
In order to have a fair comparison, we show the influence of fluctuations versus total operation time
$t_{f}$ and laser amplitude $\Omega_{0}$ on the fidelity for the STIRAP in figure 8.
As we can find, the STIRAP scheme almost perfectly restrain the influence caused by the parameters' fluctuations without doubt.
Nevertheless, in figure 7(a) we can find that the fidelity of the target
state for the STAP is still higher than $99.5\%$ even when the deviation
$\delta\Omega_{0}/\Omega_{0}=\delta t_{f}/t_{f}=10\%$,
so we can say the scheme via STAP is also robust against these variations.

Next, we will analyze the influence of dissipation induced by the atomic spontaneous
emission and the cavity decay.
The master equation of motion for the density matrix of the
whole system can be expressed as
\begin{eqnarray}\label{eq21}
\dot{\rho}&=&i[\rho,H_{I}^{\prime}]-\sum_{j=R,L}\frac{\kappa_{j}}{2}(a_{j}^{\dag}a_{j}\rho-2a_{j}\rho a_{j}^{\dag}+\rho a_{j}^{\dag}a_{j})  \cr\cr
&-&\sum_{n=1}^{3}\sum_{p=g_{0},g_{1},g_{2}}\frac{\gamma_{n,p}}{2}(\sigma_{n,p}^{\dag}\sigma_{n,p}\rho-2\sigma_{n,p}\rho\sigma_{n,p}^{\dag}+\rho\sigma_{n,p}^{\dag}\sigma_{n,p}),
\end{eqnarray}
where $\rho$ is the density operator for the whole system, $\gamma_{n,p}$ is the spontaneous emission rate from the excited
state $|e\rangle$ to the ground states $|p\rangle$ ($p=g_{0},g_{1},g_{2}$) of the $n$th ($n=1,2,3$) atom.
$\kappa_{L}$ ($\kappa_{R}$) is the decay rate of the left(right)-circular cavity mode.
For simplicity, we assume $\kappa_{L}=\kappa_{R}=\kappa$ and $\gamma_{n,p}=\gamma$.
Figure 9(a) and 9(b) shows the fidelity of the three-atom singlet state governed
by the APF Hamiltonian $H_{I}^{\prime}$ versus $\kappa/\lambda$ and $\gamma/\lambda$
with $\{$$\Omega_{0}=0.2\lambda$, $\Delta=3\lambda$, $t_{f}=40/\lambda$$\}$ and $\{$$\Omega_{0}=0.2\lambda$, $\Delta=\lambda$ and $t_{f}=40/\lambda$$\}$, respectively.
We can find the fidelity $F$ decrease slowly
with the increasing of cavity decay and atomic spontaneous emission.
When $\kappa=\gamma=0.05\lambda$,
we still can create a three-atom singlet state with a high fidelity 91.03\% as shown in figure 9(a).
By comparing figure 9(a) and (b), we find the effect of the atomic spontaneous emission
and cavity field dissipation varies with different parameters values.
So, we plot the fidelity of the three-atom singlet state versus $\kappa/\lambda$ and $\Delta/\lambda$ with
$\Omega_{0}=0.2\lambda$, $t_{f}=40/\lambda$, and $\gamma/\lambda=0$ in figure 10(a).
Figure 10(b) shows the fidelity of the three-atom singlet state versus $\gamma/\lambda$ and $\Delta/\lambda$ with
$\Omega_{0}=0.2\lambda$, $t_{f}=40/\lambda$, and $\kappa/\lambda=0$.
We find that when $\kappa/\lambda$ is nonzero, the fidelity $F$ decreases with the increasing of $\Delta/\lambda$ as shown in figure 10(a).
When $\gamma/\lambda$ is nonzero, the fidelity $F$ increases with the increasing of $\Delta/\lambda$ as shown in figure 10(b).
The phenomenon can be understood as follows.
From equation (\ref{eq19}), we know $\Omega^{\prime}\approx\sqrt{\frac{6\sqrt{2}\Delta}{t_{f}}}$,
so the laser $\Omega^{\prime}$ increases with the increasing of detuning $\Delta$.
When $\Delta$ is large enough,
the Zeno condition $\Omega^{\prime}\ll\lambda$ for the non-resonant system is not ideally fulfilled,
then the intermediate
states including the cavity-excited states would be populated
during the evolution, which would cause the system to be sensitive to the cavity decays.
In other words, as long as the detuning $\Delta$ is small, the system is robust to the cavity decays as shown in figure 10.
But substituting equation (\ref{eq19})
into the condition $\Omega^{\prime}\ll\Delta$, we deduce $\frac{6\sqrt{2}}{t_{f}}\ll\Delta$,
it denotes large $\Delta$ would be better.
So, taking the two conditions into account, when the detuning $\Delta\approx1.5\lambda$,
atomic spontaneous emission and cavity field dissipation have an equal influence in the fidelity.
According to the sensitivity of experimental apparatus to the atomic spontaneous emission and cavity field dissipation,
we can reasonably select different parameters in practical.
As we know, in general in order to restrain atomic spontaneous emission in QZD
and cavity decay in STIRAP, we introduce detuning between the atomic transition,
and that increases the evolution time.
However in our scheme we only need to select appropriate parameter
to restrain atomic spontaneous emission and cavity decay in a short time.
To sum up, it is a better choice for the experimental researchers
because the three-atom singlet state is generated much faster in
the present shortcut scheme that contributes to the experimental
research.

Finally, we present a brief discussion about the basic factors for the
experimental realization of a three-atom singlet state. In a real experiment, the cesium atoms which have been
cooled and trapped in a small optical cavity in the strong coupling
regime \cite{JYDWVHJKPrl1999,JMJRBADBAKHCPrl2003} can be used in this scheme.
The state $|e\rangle$ corresponds to $F=4, m=3$ hyperfine
state of the $6^{2}P_{1/2}$ electronic excited state, the state
$|g_{2}\rangle$ corresponds to $F=4, m=3$ hyperfine state of the
$6^{2}S_{1/2}$ electronic ground state, the state $|g_{0}\rangle$
corresponds to $F=3, m=2$ hyperfine state of the $6^{2}S_{1/2}$
electronic ground state, and the state $|g_{1}\rangle$ corresponds
to $F=3, m=4$ hyperfine state of the $6^{2}S_{1/2}$ electronic
ground state, respectively. In recent experimental conditions
\cite{SMSTJKKJKWEWHJPrA2005,MJHFGSLMBPNp06,JRBHJKPra03}, it is predicted to achieve the
parameters $\lambda=2\pi\times750$ MHz, $\kappa=2\pi\times3.5$ MHz,
and $\gamma=2\pi\times2.62$ MHz and the optical cavity mode
wavelength in a range between 630 and 850 nm. By substituting the
ratios $\kappa/\lambda=0.0047, \gamma/\lambda=0.0035$ into equation
(\ref{eq21}), we will obtain a high fidelity about 99.05\%, which shows our scheme to prepare
a three-atom singlet state is relatively robust.
Nowadays, according to the literature \cite{XWCJCDLPra14,DKJFCHMWPrL11,NBWESHJKJPWPra15,CHJAPTPHMMHCAJPrL13},
the laser pulse which is used in our scheme can be obtained in a real experiment,
so, our scheme is feasible in experiment.

\section*{Summary}
We have presented a promising method to construct shortcuts
to adiabatic passage (STAP) for a three-atom system to
generate singlet state in the cavity QED system.
We simplify a multi-qubit system
and choose the laser pulses to implement the fast generation of
entangled states in light of quantum
zeno dynamics and ``transitionless quantum driving''.
In comparison to QZD, the significant feature is that
we do not need to control the evolution time exactly.
As comparing with the STIRAP, the significant feature is
the shorter evolution time.
When dissipation is considered, we can find that the scheme is robust
against the decoherence caused by both atomic spontaneous
emission, photon leakage and operational imperfection.
In addition, the present scheme only needs to select appropriate parameter to
restrain atomic spontaneous emission and cavity decay in a short time.
Numerical simulation result shows that the scheme has a
high fidelity and may be possible to implement with the current experimental technology.
In shorts, the scheme is robust, effective and fast.
Actually, the present scheme in section \textrm{III} can be effectively applied to
$N$-atom system for preparation of $N$-atom singlet state.
We hope our work be useful
for the experimental realization of quantum information in the near future.

\section*{Acknowledgement}

  This work was supported by the National Natural Science Foundation of China under Grants No.
11575045 and No. 11374054, the Foundation of Ministry of Education
of China under Grant No. 212085, and the Major State Basic Research
Development Program of China under Grant No. 2012CB921601.

\section*{Author Contributions}

Y. X. and Z. C. came up with the initial idea for the work and
performed the simulations for the model. J. S. performed the
calculations for the model.  Z. C. and Y. H. C. performed all the
data analysis and the initial draft of the manuscript. All authors
participated in the writing and revising of the text.

\section*{Additional Information}

 Competing financial interests: The authors
declare no competing financial interests.

\newpage

\begin{figure}
 \scalebox{0.25}{\includegraphics {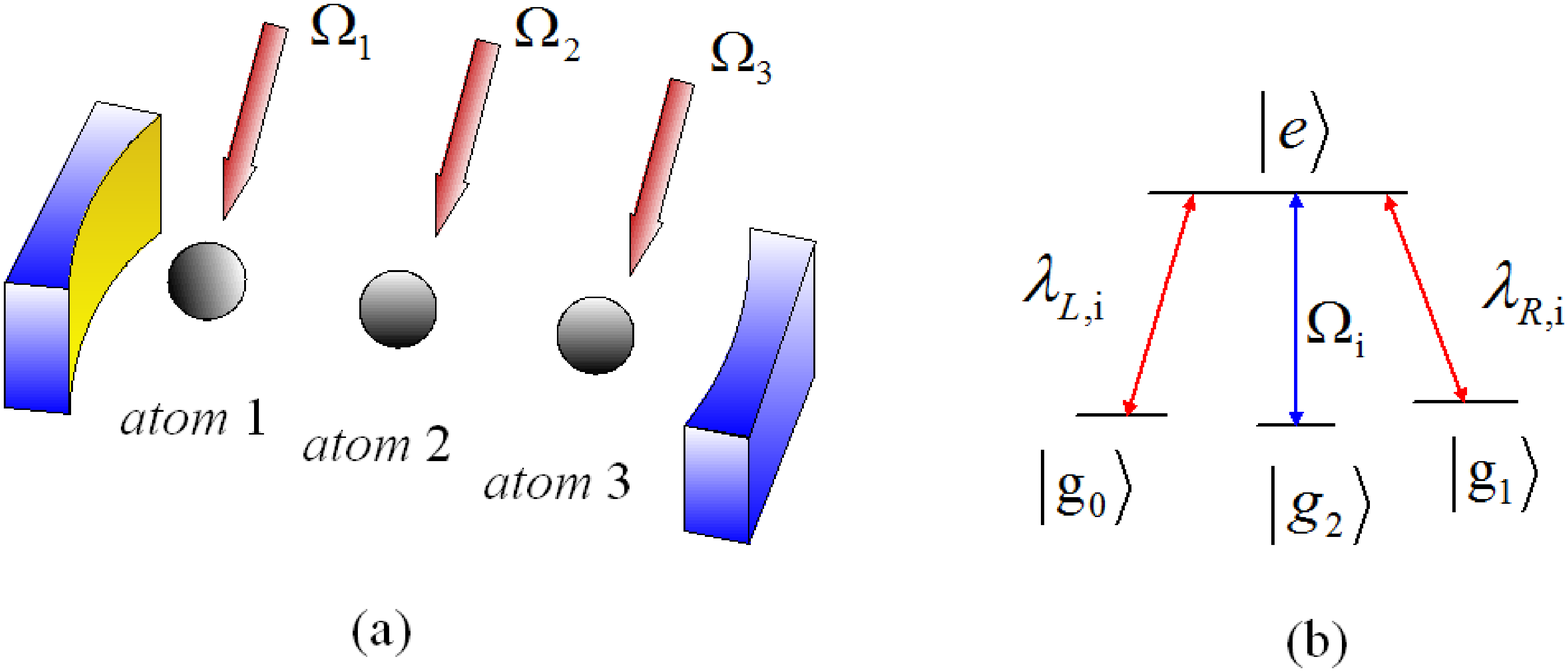}}
 \caption{
         (a) Cavity-atom combined system of the three-atom singlet state generation.
(b) Atomic level configuration for the original Hamiltonian.
         }
\end{figure}

\begin{figure}
 \scalebox{0.25}{\includegraphics {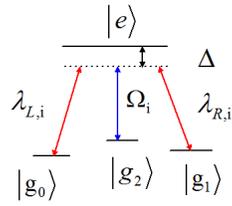}}
 \caption{
         The atomic level configuration for the APF Hamiltonian.
         }
\end{figure}
\begin{figure}
 \scalebox{0.25}{\includegraphics {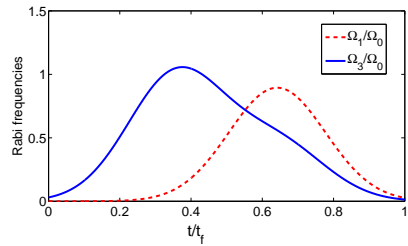}}
 \caption{
         Dependence on $t/t_{f}$ of $\Omega_{1}/\Omega_{0}$ and $\Omega_{3}/\Omega_{0}$.
         }
\end{figure}

\begin{figure}
 \scalebox{0.25}{\includegraphics {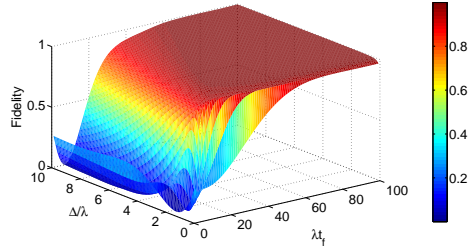}}
 \caption{
         The fidelity $F$ of the three-atom singlet state versus the interaction time $\lambda t_{f}$ and the detuning $\Delta/\lambda$.
         }
\end{figure}

\begin{figure}
 \scalebox{0.25}{\includegraphics {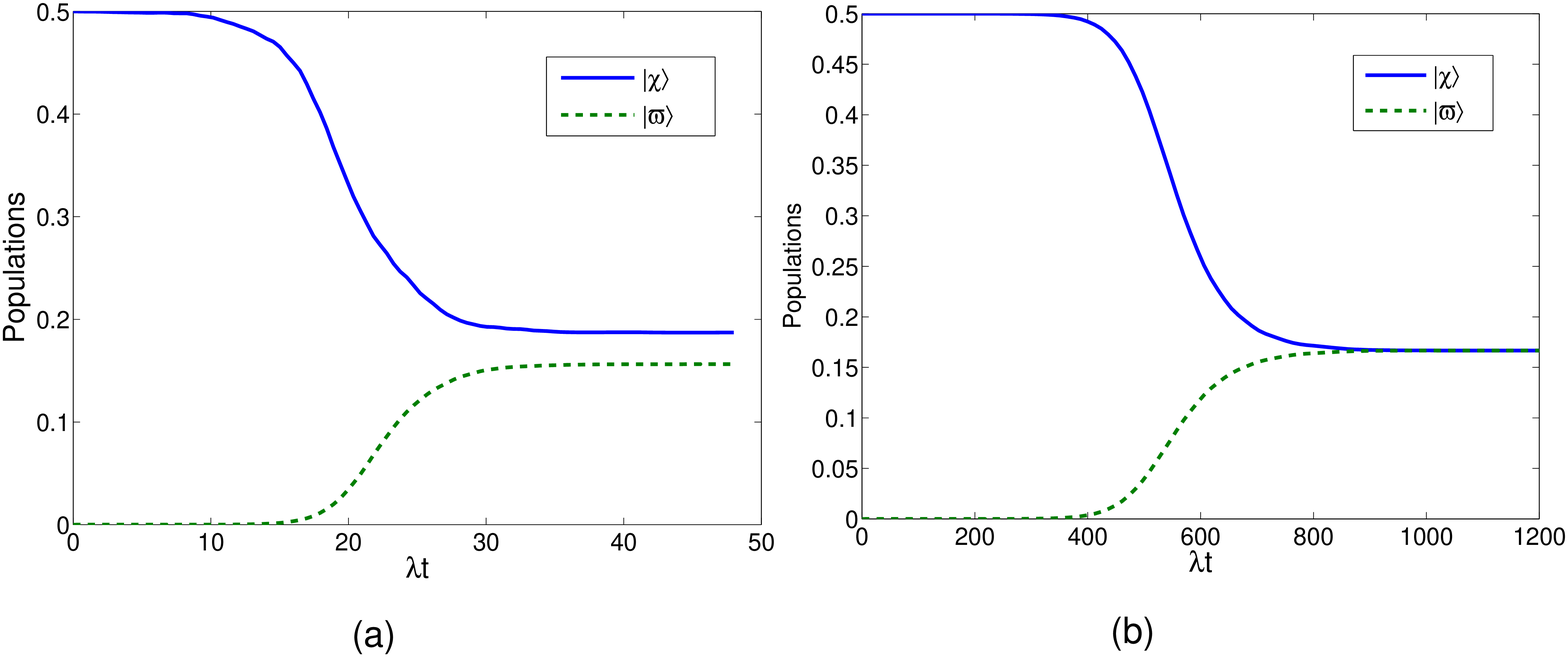}}
 \caption{
         (a) Time evolution of the populations for the states $|\chi\rangle$ and $|\varpi\rangle$ with $\Omega_{0}=0.2\lambda$, $t_{f}=40/\lambda$ and $\Delta=3\lambda$
governed by the APF Hamiltonian $H_{I}^{\prime}$. (b) Time evolution of the populations for the states $|\chi\rangle$ and $|\varpi\rangle$ with $\Omega_{0}=0.2\lambda$ and $t_{f}=1000/\lambda$
governed by the original Hamiltonian $H_{I}$.
         }
\end{figure}

\begin{figure}
 \scalebox{0.25}{\includegraphics {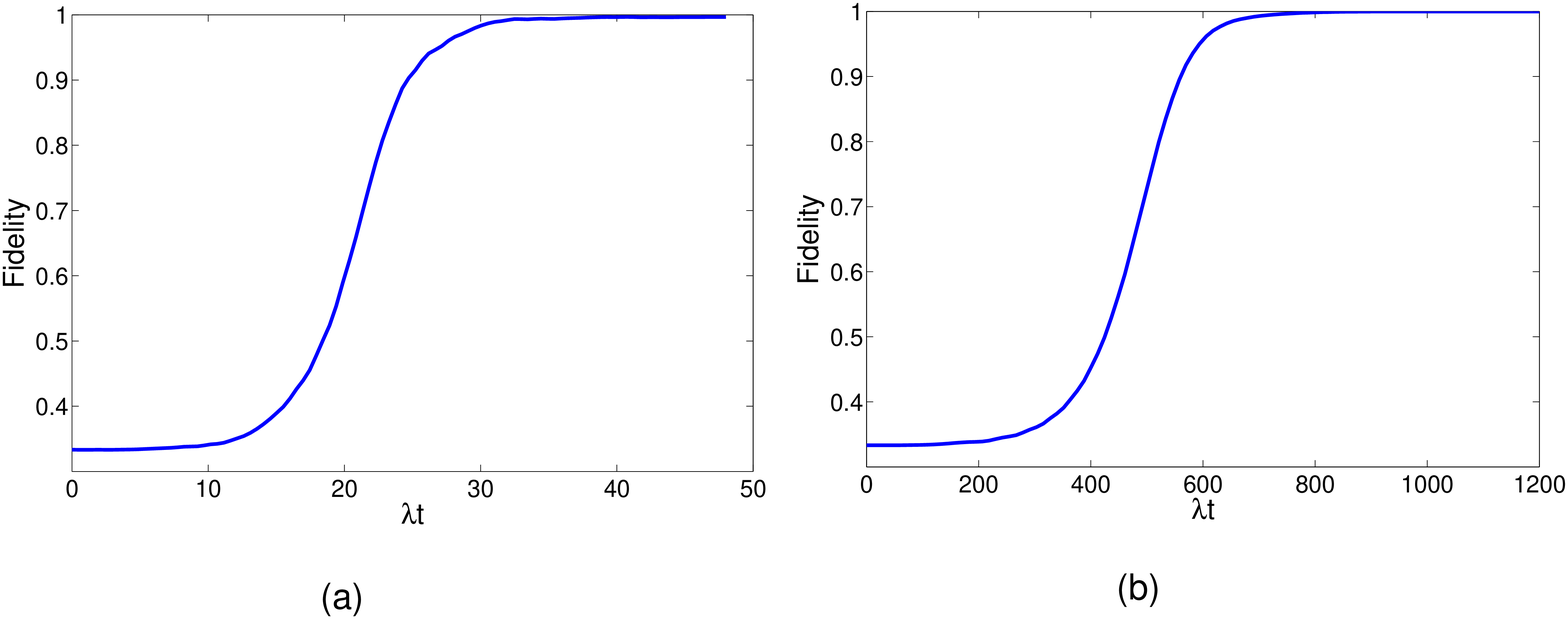}}
 \caption{
         (a) The fidelity of the three-atom single state governed by $H_{I}^{\prime}$. (b) The fidelity of the three-atom single state governed by $H_{I}$.
         }
\end{figure}

\begin{figure}
 \scalebox{0.25}{\includegraphics {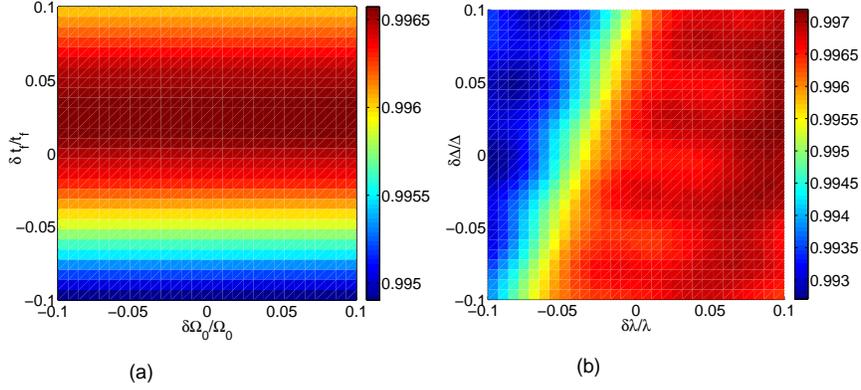}}
 \caption{
         (a) The fidelity of the three-atom singlet state versus the variations of total operation time $t_{f}$ and laser amplitude $\Omega_{0}$.
(b) The fidelity of the three-atom singlet state versus the variations of coupling constant $\lambda$ and the detuning $\Delta$.
         }
\end{figure}
\begin{figure}
 \scalebox{0.25}{\includegraphics {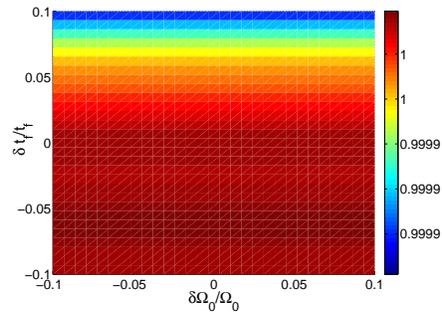}}
 \caption{
         The influence of fluctuations versus total operation time $t_{f}$ and laser amplitude $\Omega_{0}$ on the fidelity for the STIRAP.
         }
\end{figure}
\begin{figure}
 \scalebox{0.25}{\includegraphics {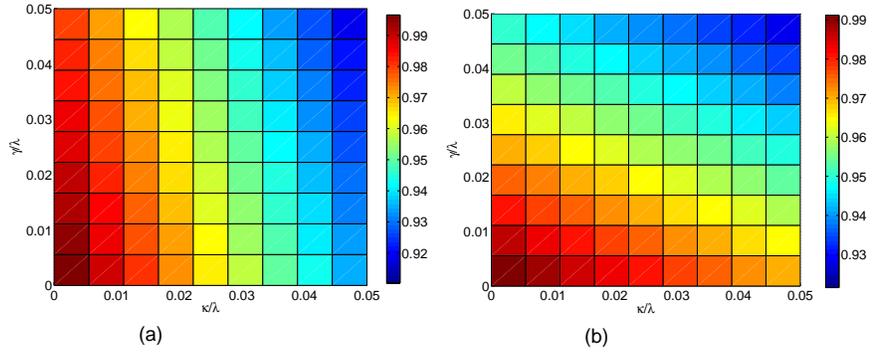}}
 \caption{
         (a) The fidelity of the three-atom singlet state governed by the APF Hamiltonian $H_{I}^{\prime}$ versus $\kappa/\lambda$ and $\gamma/\lambda$
with $\Omega_{0}=0.2\lambda$, $\Delta=3\lambda$ and $t_{f}=40/\lambda$. (b) The fidelity of the three-atom singlet state governed by the APF Hamiltonian $H_{I}^{\prime}$ versus $\kappa/\lambda$ and $\gamma/\lambda$ with $\Omega_{0}=0.2\lambda$, $\Delta=\lambda$ and $t_{f}=40/\lambda$.
         }
\end{figure}
\begin{figure}
 \scalebox{0.25}{\includegraphics {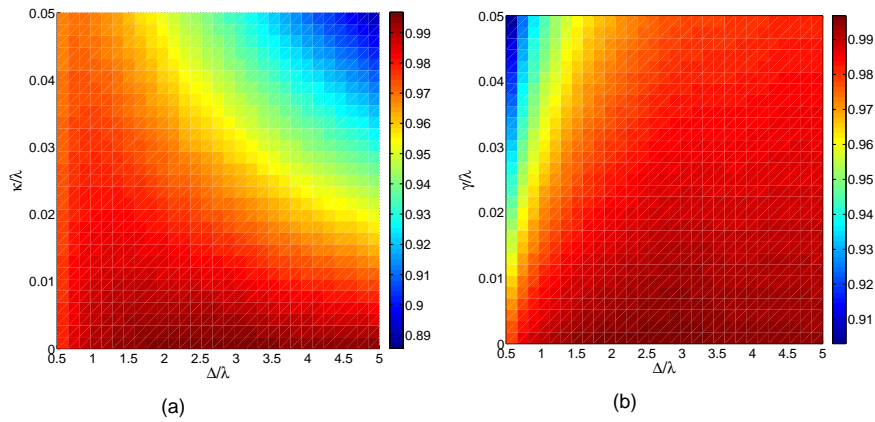}}
 \caption{
         (a) The fidelity of the three-atom singlet state versus $\kappa/\lambda$ and $\Delta/\lambda$ with
$\Omega_{0}=0.2\lambda$, $t_{f}=40/\lambda$, and $\gamma/\lambda=0$. (b) The fidelity of the three-atom singlet state versus $\gamma/\lambda$ and $\Delta/\lambda$ with
$\Omega_{0}=0.2\lambda$, $t_{f}=40/\lambda$, and $\kappa/\lambda=0$.
         }
\end{figure}

\end{document}